\documentclass[twocolumn,showpacs,preprintnumbers,amsmath,amssymb]{revtex4}
\usepackage{graphicx}
\usepackage{dcolumn}
\usepackage{bm}
\usepackage{latexsym}
\usepackage{epsfig}
\begin{document}
\preprint{APS/123-QED}
\title{Elastic proton-deuteron scattering at intermediate energies}
\author{\firstname{A.}~\surname{Ramazani-Moghaddam-Arani$^{1,2}$}}
\hspace{-5cm}\email{ramazani@kvi.nl}
\author{\firstname{H.R.}~\surname{Amir-Ahmadi$^{1}$}}
\author{\firstname{A.D.}~\surname{Bacher$^{3}$}}
\author{\firstname{C.D.}~\surname{Bailey$^{3}$}}
\author{\firstname{A.}~\surname{Biegun$^{1}$}}
\author{\firstname{M.}~\surname{Eslami-Kalantari$^{1}$}}
\author{\firstname{I.}~\surname{Ga\v{s}pari\'c$^{4}$}}
\author{\firstname{L.}~\surname{Joulaeizadeh$^{1}$}} 
\author{\firstname{N.}~\surname{Kalantar-Nayestanaki$^{1}$}}
\author{\firstname{St.}~\surname{Kistryn$^{5}$}}
\author{\firstname{A.}~\surname{Kozela$^{6}$}}
\author{\firstname{H.}~\surname{Mardanpour$^{1}$}}
\author{\firstname{J.G.}~\surname{Messchendorp$^{1}$}}\email{messchendorp@kvi.nl}
\author{\firstname{A.M.}~\surname{Micherdzinska$^{7}$}}
\author{\firstname{H.}~\surname{Moeini$^{1}$}}
\author{\firstname{S.V.}~\surname{Shende$^{1}$}}
\author{\firstname{E.}~\surname{Stephan$^{8}$}}
\author{\firstname{E.J.}~\surname{Stephenson$^{3}$}}
\author{\firstname{R.}~\surname{Sworst$^{5}$}}
\affiliation{%
$^1$ KVI, University of Groningen, Groningen, The Netherlands}
\affiliation{%
$^2$ Department of Physics, Faculty of Science, University of Kashan, Kashan, Iran}
\affiliation{%
$^3$ Indiana University Cyclotron Facility, Indiana, USA}%
\affiliation{%
$^4$ Rudjer Bo\v{s}kovi\'c Institute, Zagreb, Croatia}
\affiliation{%
$^5$ Institute of Physics, Jagiellonian University, Cracow, Poland}
\affiliation{%
$^6$ Henryk Niewodnicza\'nski, Institute of Nuclear Physics, Cracow, Poland}
\affiliation{%
$^7$ University of Winnipeg, Winnipeg, Canada}%
\affiliation{%
$^8$ Institute of Physics, University of Silesia, Katowice, Poland}%
 
\date{\today}
\begin{abstract}
Observables in elastic proton-deuteron scattering are
sensitive probes of the nucleon-nucleon interaction and
three-nucleon force effects. The present experimental data base for
this reaction is large, but contains a large discrepancy between
data sets for the differential cross section taken at 135~MeV/nucleon
by two experimental research groups. This paper reviews the background
of this problem and presents new data taken at KVI. Differential cross
sections and analyzing powers for the $^{2}{\rm H}(\vec p,d){p}$ and
${\rm H}(\vec d,d){p}$ reactions at 135~MeV/nucleon and
65~MeV/nucleon, respectively, have been measured.  The data differ
significantly from previous measurements and consistently follow the
energy dependence as expected from an interpolation of published data
taken over a large range at intermediate energies.
\end{abstract}
\maketitle
The nucleon-nucleon potential (NNP) has been studied extensively by
investigating the properties of bound nuclear systems, and, in more
detail, via a comparison of high-precision two-nucleon scattering data
with modern potentials based on the exchange of
bosons~{\cite{bon87,argon,nij1}}. A few of the modern NNPs were
facilitated by a partial-wave analysis (PWA), that provides a
nearly model-independent analysis of the available scattering
data~{\cite{nij2}}. The modern NNPs reproduce the world data base with
a reduced chi-square close to one and have, therefore, been accepted
as high-quality benchmark potentials.  The precision of modern NNPs
has given confidence to study in detail the three-nucleon potential
(3NP) which was already predicted in 1939 by Primakoff and
Wilson~{\cite{prima}}. Compelling evidence of 3NP effects came from
various recent theoretical and experimental studies. For example, for
light nuclei, Green's function Monte-Carlo calculations employing the
high-quality NNPs clearly underestimate the experimental binding
energies~{\cite{argon}}, and, therefore, show that NNPs are not
sufficient to describe the three-nucleon and heavier systems
accurately. In the last decade, high-precision data at intermediate
energies in elastic \it {Nd} \rm and \it {dN} \rm
scattering~{\cite{kars01,kars03,kars05,bieber,sakai00,kimiko05,kimiko02,postma,hamid,kurodo,mermod,Igo,ald,Hos,Ela07,shimi,hatan,IUCF}}
for a large energy range together with rigorous Faddeev
calculations~{\cite{gloeckle}} for the three-nucleon system have
proven to be a sensitive tool to study the 3NP. In particular, a large
sensitivity to 3NP effects exists in the minimum of the differential
cross section~{\cite{witala98,nemoto}}. Precision data for a large
energy interval for the differential cross section and analyzing power
came from recent experimental studies at KVI
~{\cite{kars01,kars03,kars05}}, RIKEN ~{\cite{sakai00}} and
RCNP~{\cite{{kimiko05}}. All these experiments had one common energy
of 135~MeV/nucleon. Strikingly, the cross sections obtained at KVI
were found to be significantly larger than those measured at RIKEN and
at RCNP. The KVI data show significant deviation from predictions of
state-of-the-art Faddeev calculations incorporating modern NNPs and
3NPs at this energy, whereas the results obtained at RIKEN and RCNP
imply that the cross section can be described reasonably well
exploiting the same potentials.\\
\indent This paper presents the results of a
new measurement of the differential cross sections of the reaction
$^{2}{\rm H}(\vec p,d){p}$ at a proton-beam energy of 135~MeV, taken
to provide additional data at 135 MeV/nucleon. These results are
compared with the previously published data taken at intermediate
energies. The data are obtained at KVI using a new experimental
equipment, Big Instrument for Nuclear-polarization Analysis
(BINA). Also, systematic uncertainties were checked by measuring the inverse
reaction ${\rm H}(\vec d,d){p}$ at a deuteron-beam energy of
65~MeV/nucleon using the same experimental setup and analysis
methods. The differential cross section at this energy is
well-known~\cite{shimi} and can, therefore, be exploited to verify
independently the read-out and analysis procedure, the applied
detector inefficiencies, and the beam-current measurement.
For both experiments, polarization observables have also been measured.
This allows to check some of the aspects
of the experiment and the analysis procedure, excluding the absolute normalization,  
by comparing our results with data from the literature. Important to note is
that the previously published KVI data were obtained using the
Big-Bite Spectrometer, which was located at a different beam
line and which used a completely different analysis framework, target,
read-out and data-acquisition system.

BINA is a setup with a nearly 4$\pi$ geometrical acceptance and has
been used in various few-nucleon scattering experiments to measure the
scattering angles and energies of protons and deuterons with the
possibility for particle identification. The detector is composed of a
forward and a backward part.
%
The forward part consists of a Multi-Wire Proportional Chamber (MWPC)~\cite{Vol99}
and a segmented hodoscope of vertically-placed thin scintillators with
a thickness of 2~mm followed by ten horizontally-placed scintillators
with a thickness of 12~cm each. The thick scintillators were mounted
in a cylindrical shape, thereby, pointing to the target. The thickness
of these scintillators is sufficient to stop all the protons and
deuterons originating from the processes described in this paper.
 The detection efficiency of the MWPC was obtained by using
an unbiased and nearly background-free data sample of
elastically-scattered deuterons and was found to be
typically~97$\pm$1\%.  The backward part has 149 phoswich
scintillators covering polar angles between 40$^\circ$ to 165$^\circ$
with a nearly-full azimuthal coverage.

The polarized proton beams from the AGOR accelerator impinged on a
liquid-deuterium~\cite{Nasser} target with a thickness of 3.85~mm, with an
uncertainty of 5\%, which was mounted in the center of the backward
part of BINA. The target cell is made of high-purity aluminum to
optimize the thermal conductivity and the windows are covered by a
transparent foil of Aramid with a thickness of 4~$\mu$m. The cell was
operating under a pressure of 250~mbar and a temperature of 19~K. For
the deuteron beam, a solid organic CH$_2$ target was used for the
${\rm H}(\vec d,d){p}$ experiment with an effective thickness of
13.75$\pm$0.24 ~mg/cm$^2$.  The beam current was typically 15~pA and
was monitored continuously during the experiment via a Faraday cup at
the end of the beam line. The current meter was calibrated using a
precision current source with an uncertainty of about 2\%.\\
\hspace{-1cm}\begin{figure}[t]
 \centering 
\epsfxsize=9.8cm
\epsfbox{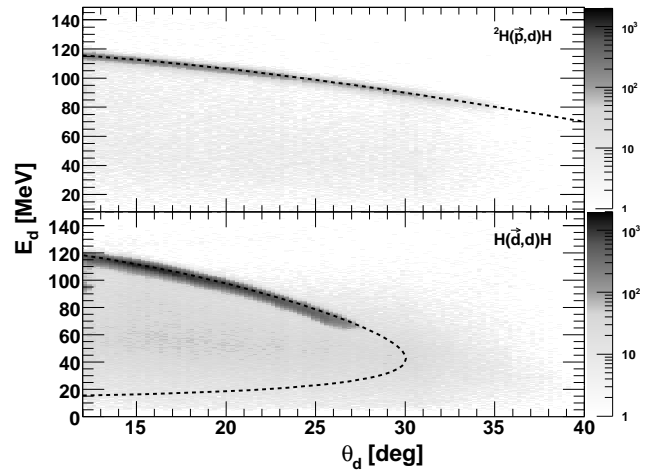} 
\vspace*{-0.5cm}
 \caption{The figure depicts the correlation between the scattering
 angle and the deposited energy for events registered in the forward
 part of BINA. The top panel corresponds to data taken using a proton
 beam with an energy of 135~MeV impinging on a liquid-deuterium
 target. The bottom panel shows data taken with a deuteron beam of
 130~MeV impinging on a solid CH$_2$ target. The dashed lines
 represent the expected kinematical correlation for a scattered
 deuteron in the elastic proton-deuteron reaction.}\label{loci}
\end{figure}
\begin{figure*}[ht]
\centering
\resizebox{14cm}{!}{\includegraphics[angle = 0,width =1\textwidth]{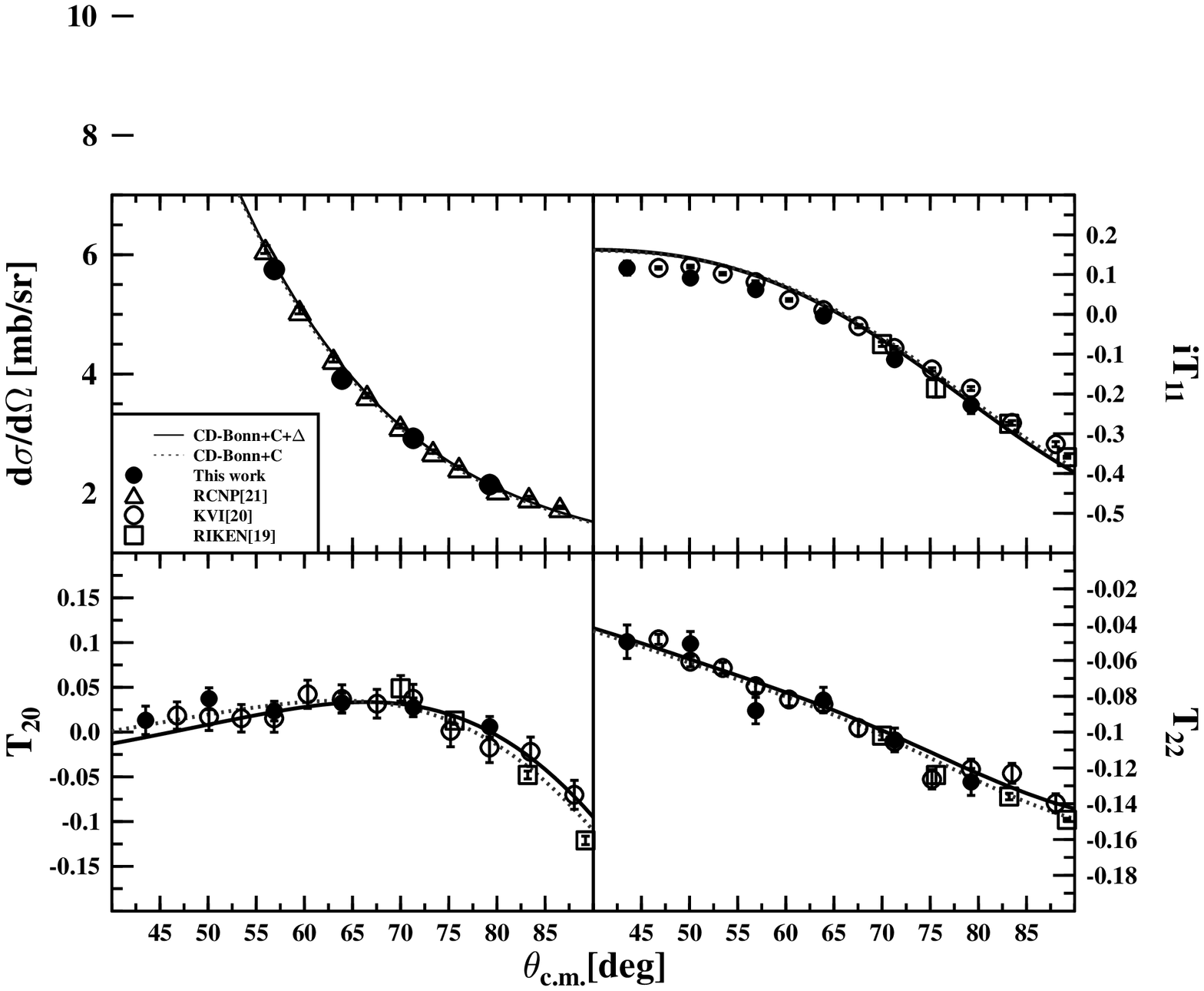}}
\vspace*{-0.4cm}
\caption{The differential cross section, vector and tensor analyzing
 powers of the elastic $\vec d+p $ reaction at $E_{d}^{lab}$=130~MeV
 as a function of the center-of-mass angle, $\theta_{c.m.}$. In each
 panel, filled circles are data from this work.  Only statistical
 uncertainties are shown for each point. The open triangles in
 top-left panel represent cross section data taken at
 RCNP~\cite{shimi}. The open circles and open squares in the other
 panels are analyzing power data taken at KVI~\cite{Ela07} and at
 RIKEN~\cite{Hos}, respectively. Note that we do not present differential 
 cross sections for $\theta_{cm}<55^{\circ}$. At these angles, an increasing 
 fraction of backward-scattered protons are undetected due to their 
 small energy and the amount of material between the target and the
 detector. The solid lines represent the results of a coupled-channel
 calculation by the Hannover-Lisbon theory group and are based on the
 CD-Bonn potential including the Coulomb interaction and an
 intermediate $\Delta$-isobar. The results of a similar calculation,
 however, excluding the $\Delta$-isobar, are shown as dashed lines.}
\label{dpresults}
\end{figure*}
For this experiment, events were selected in which deuterons were
detected in the forward part in coincidence with a scattered proton in
the backward part of BINA. The efficiency of the coincidence hardware
trigger was determined from a data sample obtained from a minimum-bias
trigger and found to be 98$\pm$1\%. In the analysis, a large part of
the background, pre-dominantly from the break-up reactions, was
reduced by verifying that the forward and backward particles
determined the same scattering plane within a limit of $\pm
20^\circ$. Figure~\ref{loci} shows the correlation between the
scattering angle of particles and their deposited energy in the
forward part of BINA for both incident energies.
 The reactions $^{2}{\rm H}(\vec p,d){p}$ and ${\rm H}(\vec d,d){p}$
 can clearly be identified in the top and bottom panel, respectively.
 It can be seen that for both reactions, the forward-scattered
 deuterons have similar energies; thus the check against lower-energy
 data is relevant at 135 MeV/nucleon.  Peak sums were obtained by
 transforming the forward energy spectra to the center-of-mass system,
 reproducing these spectra with the sum of a Gaussian and a polynomial
 function of energy, and using for the peak sum the integral of the
 Gaussian function.\\
%
%
%
%
%
\indent Part of the events away from the loci in Fig.~\ref{loci} corresponds
to elastically-scattered deuterons which undergo a hadronic
interaction inside the scintillators. These events cannot be separated
easily from the background from the break-up reaction. The contribution
of particles undergoing hadronic interactions was obtained by
analyzing part of the data for which stringent cuts on the scattering
angle and energy of the backward scattered particles were applied.
 Since these cuts removed essentially all break-up
background, events which fell outside the full energy peak were
interpreted as the fraction of the elastic events that comprised the
hadronic interaction tail below the range of integration of the
Gaussian peak. The total fraction of events which suffer from a
hadronic interaction is about 16\% with an uncertainty of 2\%.\\
 \begin{figure}[h]
 \centering 
\epsfxsize=8.5cm\epsfysize=10.cm\epsfbox{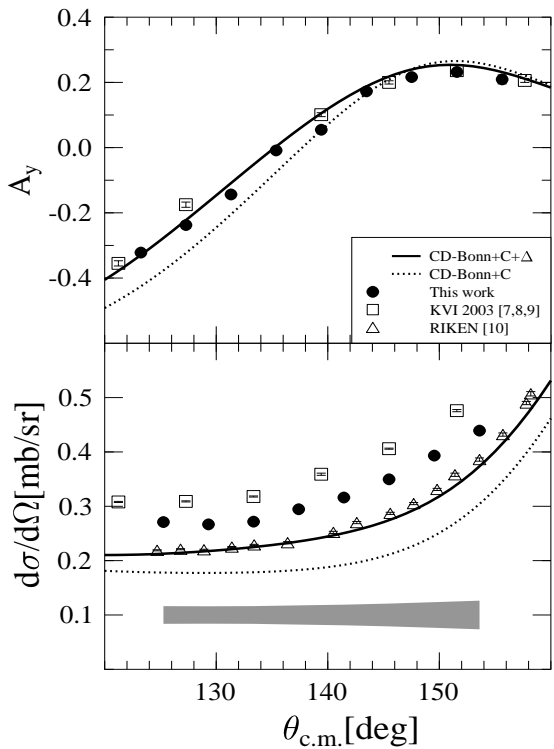} 
 \caption{The top panel shows the results for the analyzing power of
 the $^{2}{\rm H}(\vec p,d){p}$ reaction. The present work is
 represented by filled circles and the previously-published KVI
 data~\cite{kars01} are shown by open squares. The bottom panel shows
 the corresponding measurement of the differential cross section. Our
 results are represented by filled circles and are compared to data
 published in Refs.~\cite{sakai00,kars03}. The error bars represent
 statistical uncertainties. The width of the gray band corresponds to the systematic
 uncertainty of our measurement. For a
 description of the lines, we refer to the caption of
 Fig.~\ref{dpresults}.}
\label{cros}
\vspace{-7mm}
\end{figure}
\indent Although the emphasis of this work is to obtain differential cross
sections in elastic proton-deuteron scattering, we have measured the
various polarization observables for these reactions as well. A
comparison with the existing and well-established world data base of
vector and tensor analyzing powers allows us to verify the quality of
the data. Beams of vector-polarized protons or vector and
tensor-polarized deuterons were produced in an atomic-beam-type ion
source. The polarization of the beams was measured using a
Lamb-shift polarimeter (LSP) in the low-energy beam line and by an
in-beam polarimeter (IBP) which was installed at the high-energy beam
line after acceleration.
More information on the operation of the IBP and LSP can be found in
Refs.~\cite{ibp,lsp}. During the experiments discussed in this paper,
the beam polarization was typically 65\% of the maximally-allowed
theoretical values. The analyzing powers of the elastic $\vec d+p$ and
$\vec p+d$ reactions were obtained by studying the dependence of the
differential cross section on the azimuthal scattering angle of the
deuterons in the final state. An equivalent data analysis has been
reported and published in Ref.~\cite{Ela07}.\\
\indent Figure~\ref{dpresults} summarizes the measured cross sections and
analyzing powers with only statistical errors for the ${\rm H}(\vec
d,d){p}$ reaction at a beam energy of 65~MeV/nucleon and as a function
of the center-of-mass angle, $\theta_{c.m.}$.  The systematic
uncertainty of the measured differential cross section and analyzing
powers is estimated to be 4\% and 3\%, respectively. The data of this
experiment are shown as filled circles and are compared with published
cross section data taken in the past at KVI and elsewhere.
 The new results are in good agreement with the previously published
 data and, thereby, demonstrate that the experimental setup and the
 data-analysis procedure are well understood. The solid
 lines in Fig.~\ref{dpresults} are the result of a rigorous Faddeev
 calculation by the Hannover-Lisbon theory group. 
 In this calculation, the $\Delta$-isobar excitation mediates an
 effective 3NP with prominent Fujita-Miyazawa and Illinois ring-type
 contributions. These contributions are based on the exchange of
 $\pi$, $\rho$, $\omega$, and $\sigma$ mesons. For the dashed lines,
 the $\Delta$-isobar degree of freedom was switched off. The
 difference between the solid and dashed lines represents, therefore,
 a three-nucleon force effect within the framework of this model. Note
 that at this energy, the data are very well described by the
 state-of-the art calculations.\\ \indent Figure~\ref{cros} shows our
 results (filled circles) for the vector analyzing power (top panel)
 and the differential cross section (bottom panel) of the ${\rm
 ^2H}(\vec p,d){p}$ reaction at a beam energy of 135~MeV and as a
 function of the center-of-mass angle, $\theta_{c.m.}$. The error bars
 for all symbols represent statistical uncertainties, which, in most
 cases, are smaller than the symbol sizes. The measured analyzing
 powers, shown in the top panel of Fig.~\ref{cros}, are compared to
 previously-published KVI data from Ref.~\cite{kars01} (open
 squares). The two data sets are in very good agreement and the
 comparison gives confidence in the quality of our data apart from an
 overall normalization. Note, however, the striking discrepancy
 between the previously-published differential cross sections measured
 at KVI~\cite{kars03} (open squares) and RIKEN~\cite{sakai00} (open
 triangles) as shown in the bottom panel of Fig.~\ref{cros}. The
 systematic uncertainties of the cross section measurements are not
 shown in the figure and are 2\% for the RIKEN data, 4.5\% for the
 previously-published KVI data, and 6\% for the data discussed in this
 paper. Our results fall between these two data sets and differ
 significantly from both of the previous measurements. The solid and
 dashed lines represent the results of the Faddeev calculations by the
 Hannover-Lisbon theory group. The RIKEN data set would imply that the
 theoretical description of this three-nucleon process is well
 understood  whereas, the KVI data indicate that not all the
 ingredients are incorporated yet in the models. The observed
 discrepancy between data and theory could originate from the fact
 that the predictions are solutions of the non-relativistic Lippmann
 Schwinger equations, which might  imply that not all relativistic
 effects are included consistently.\\
\begin{figure}[h]
\centering 
\epsfxsize=9.5cm\epsfysize=8.5cm\epsfbox{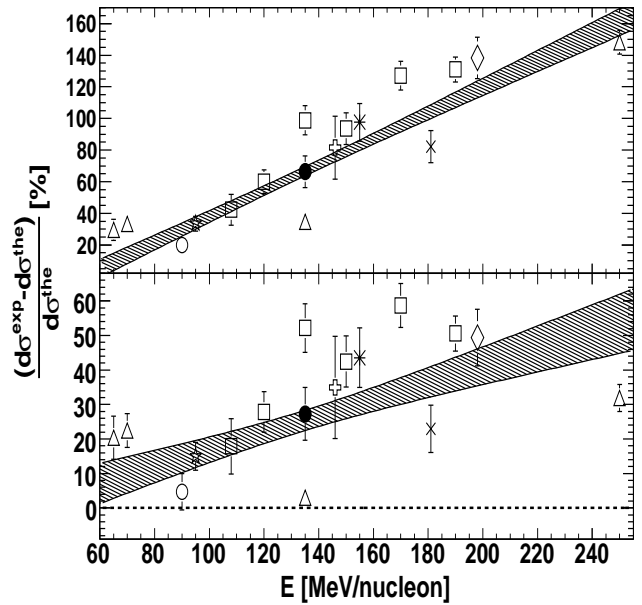}
\caption{The relative difference between the calculations by the Hannover-Lisbon theory group and the
 measured cross sections for the elastic $p+d$ reaction as a
 function of beam energy for $\theta_{c.m.}=140 ^\circ$. 
 The top panel shows the differences with a calculation based
 on the CD-Bonn potential and the Coulomb interaction, whereas for the 
 bottom panel also an additional $\Delta$ isobar has been taken into account.
 Open squares are data from~{\cite{kars03}}, open triangles are data from
 Refs.~\cite{sakai00,kimiko02,shimi,hatan}, open circle is from~{\cite{hamid}},
 open star is from~{\cite{mermod}}, cross is from~{\cite{Igo}}, star
 is from~{\cite{kurodo}}, open cross is from~{\cite{postma}}, diamond is from
 ~{\cite{ald}} and the filled circle is from this work.  The shaded
 band represents the result of a line fit through
 the data excluding the results obtained at KVI, RIKEN and RCNP. 
 The width of the band corresponds to a 2$\sigma$ error of the fit.}
\label{syscheck} 
\end{figure}
 The observed inconsistency for the differential cross section, as
 shown in Fig.~\ref{cros}, initiated a discussion within the
 nuclear-physics community on the reliability of the experimental data
 and on how to interpret the data in terms of underlying physics, such
 as 3NP effects. It is, therefore, of importance to review these
 observations with respect to three-nucleon scattering data taken at
 other energies and in other channels. The most generic approach would
 be to perform a partial-wave analysis (PWA) of all available
 three-nucleon scattering data similar to what has been done for the
 nucleon-nucleon scattering data by the Nij\-me\-gen
 group~{\cite{nij2}}. Such an analysis would provide an independent
 judgment of the quality of the data sets. At present, a PWA in the
 three-nucleon sector is still a technical challenge which, so-far,
 has not been pursued and which is outside the scope of this
 paper. Instead, we have carried out a systematic study of the energy
 dependence of all available cross sections in elastic proton-deuteron
 scattering with respect to state-of-the-art calculations by the
 Hannover-Lisbon theory group. One example from this study is
 presented in Fig.~\ref{syscheck}. The top panel shows the relative
 difference between the model predictions excluding the $\Delta$
 isobar contribution and data taken at a fixed center-of-mass angle of
 $\theta_{c.m.}$=140$^\circ$. The data points were extracted from a
 polynomial fit through each angular distributions. The error
 bars correspond to a quadratic sum of the statistical and systematic
 uncertainties of each measurement. We have made a straight-line fit
 through the data, excluding the measurements performed at KVI, RIKEN,
 and RCNP. Including these data in the fit will, however, not change 
 the conclusions. The shaded band shows the fit function
 including its errors, which are calculated based on the correlation
 matrix of the fit parameters. Note the rapidly increasing discrepancy
 with energy between data and the predictions based on a high-quality
 two-nucleon potential. At 135~MeV/nucleon the deviation is already
 more than 50\%, which is supported by our data. Based on the RIKEN
 data one would conclude that the large deviation can be resolved at
 135~MeV/nucleon by including the $\Delta$-isobar effect, as
 demonstrated in the bottom panel of Fig.~\ref{syscheck}. However, if
 this is true, it would imply that almost all experiments must suffer
 from a normalization problem. A similar study has been done for other
 center-of-mass angles, $\theta_{c.m.}$, between 125$^\circ$ and
 155$^\circ$, leading to the same observation and conclusion. \\
\indent In conclusion, we have analyzed the observed discrepancy between two
experimental data sets of the differential cross sections in elastic
$\vec p+d$ scattering at 135~MeV, which were obtained in the past at
RIKEN and at KVI. For this, we have remeasured the differential cross
section at center-of-mass angles between 125$^\circ$ and 155$^\circ$, and compared our
results with previously-published data. Furthermore, the differential
cross section and analyzing powers of the inverse reaction, $\vec d+p$
at a beam-energy of 65~MeV/nucleon have been obtained and compared to
the existing and well-established data-base at these relatively low
energies. The data for this reaction are found to be in excellent
agreement with the existing data base, which proves that BINA and our
analysis procedure are well suited to measure the elastic channel with
high precision.  The measured differential cross section for the
elastic reaction at 135~MeV differs significantly from the
previously-published data taken at RIKEN and at KVI. We have carried
out a systematic study of the energy dependence of all available cross
sections in elastic proton-deuteron scattering. The results of this
study revealed that our data consistently follow the interpolated
energy dependence, whereas, in particular, the RIKEN data deviate
significantly from the expected trend.\\
\indent The authors acknowledge the work by the cyclotron and ion-source
groups at KVI for delivering high-quality polarized beams.
We thank Frans Mul for the mechanical
construction of BINA and the Hannover-Lisbon group for providing the
theoretical calculations. This work is part of the research program of
the ``Stichting voor Fundamenteel Onderzoek der Materie'' (FOM) with
financial support from the ``Nederlandse Organisatie voor
Wetenschappelijk Onderzoek'' (NWO).  Furthermore, the present work has
been performed with financial support from the University of Groningen
and the Gesellschaft f\"ur Schwerionenforschung mbH (GSI),
Darmstadt. The Indiana co-authors were supported by US NFS grant PHY-0457219.\\
%

\end{document}